# Reinforcement Learning for Optimizing Large Qubit Array based Quantum Sensor Circuits

Laxmisha Ashok Attisara
*Department of Electrical Engineering and Computer Science*
*Cleveland State University*
Cleveland OH, 44115, U.S.A.
l.ashokattisara@vikes.csuohio.edu

Sathish A. P. Kumar
*Department of Electrical Engineering and Computer Science*
*Cleveland State University*
Cleveland OH, 44115, U.S.A.
s.kumar13@csuohio.edu

*Abstract* -- **As the number of qubits in a sensor increases, the complexity of designing and controlling the quantum circuits grows exponentially. Manually optimizing these circuits becomes infeasible. Optimizing entanglement distribution in large-scale quantum circuits is critical for enhancing the sensitivity and efficiency of quantum sensors [5], [6]. This paper presents an engineering integration of reinforcement learning with tensor-network-based simulation (MPS) for scalable circuit optimization for optimizing quantum sensor circuits with up to 60 qubits. To enable efficient simulation and scalability, we adopt tensor network methods—specifically, the Matrix Product State (MPS) representation—instead of traditional state vector or density matrix approaches. Our reinforcement learning agent learns to restructure circuits to maximize Quantum Fisher Information (QFI) and entanglement entropy while reducing gate counts and circuit depth. Experimental results show consistent improvements, with QFI values approaching 1, entanglement entropy in the 0.8–1.0 range, and up to 90% reduction in depth and gate count. These results highlight the potential of combining quantum machine learning and tensor networks to optimize complex quantum circuits under realistic constraints.**

## I. INTRODUCTION AND BACKGROUND

Quantum sensing relies on entanglement to enhance measurement precision, making entanglement distribution a crucial design factor in sensor circuits[2]. While prior research has explored optimization for small-scale quantum systems, scaling up to larger circuits remains challenging due to the exponential growth in computational complexity[1]. Traditional simulations using state vector or density matrix formalism become computationally infeasible for circuits beyond 15-20 qubits. To overcome this, our work shifts to a tensor network-based simulation framework, utilizing Matrix Product States (MPS) to efficiently represent and manipulate quantum states. This enables scalable training and evaluation of circuits up to 60 qubits.

We propose a reinforcement learning (RL) approach that dynamically modifies circuit structures to maximize sensitivity, as quantified by Quantum Fisher Information (QFI), and maintain strong entanglement, measured by entanglement entropy. Our multi-reward optimization framework also minimizes gate counts and circuit depth—key factors in reducing noise and improving implement ability on quantum hardware.

This paper introduces a scalable, noise-aware, and adaptive framework combining MPS-based simulation with deep RL optimization to push the limits of quantum sensor circuit design using sophisticated optimization techniques like entanglement injection, boosting, adaptive learning, entanglement layers. While our method leverages existing tools such as DDQN, QFI, and tensor network methods (MPS), the novelty lies in their integration into a unified framework specifically targeting scalability for quantum sensor circuits. Rather than introducing a fundamentally new RL algorithm, this work contributes by engineering a practical, noise-aware, and scalable pipeline for entanglement optimization.

## II. RESEARCH OBJECTIVE

The primary objective of this research is to develop scalable algorithms to simulate deep quantum circuits involving medium-larger sized arrays of qubits efficiently by reducing the complexity under noise environment. Quantum Sensor network uses the developed QML Algorithms to simulate and optimize the layout, distribute and maintain entanglement efficiently.

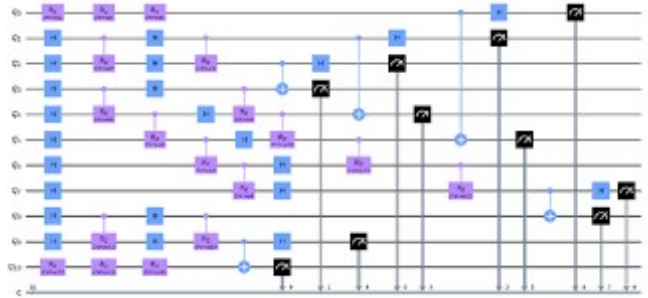

Fig. 1. Entanglement distribution in a larger quantum circuit.

## III. RELATED WORKS

A. Quantum circuit synthesis and transpiling with Reinforcement Learning work demonstrates the integration of Reinforcement Learning (RL) into quantum transpiling workflows, significantly enhancing the synthesis and routing of quantum circuits. By employing RL, they achieve near-optimal synthesis of Linear Function, Clifford, and Permutation circuits, up to 9, 11 and 65 qubits respectively. Though the goal is synthesis of larger quantum circuits they have focused more on transpilers and Clifford circuits which are not helpful in quantum sensing.[5]

B. Quantum Circuit Optimization with Deep RL. This is an approach to quantum circuit optimization based on RL. It demonstrates the feasibility of approach by training agents on 12 qubits random entangled circuit where on an average depth reduction by 27% and CNOT gate counts by 15%. Although the approach concentrates on 2 qubits gate optimization, the amount of entanglement is not measured during the process.[6]

C. Reinforcement Learning based Quantum circuit optimization via ZX-Calculus. It proposes a method for optimizing quantum circuits using graph-theoretic simplification rules of ZX-diagrams. It illustrates its versatility by targeting both total and two-qubit gate count reduction, conveying the potential of tailoring its reward function to the specific characteristics. Here the work is better compared to other ZX calculus and heuristic algorithms but remain competitive in terms of computational performance. It does not focus on the sensitivity while optimizing the circuits .[7]

Recent frameworks such as circ-RL, ZX-Calculus-based optimizers [7], Maslov's depth optimization [12], and Qiu et al.'s entanglement generation for metrology [13] highlight alternative approaches. Our framework differs by directly combining RL with tensor network simulation for scalable entanglement-aware optimization, enabling evaluation up to 60 qubits, a regime not addressed by prior work.

## IV. METHODOLOGY

We propose a scalable framework for optimizing quantum sensor circuits using a custom Deep Reinforcement Learning (RL) pipeline integrated with Tensor network backend. Our approach targets circuits up to 60 qubits and 120 gates, maintaining high Quantum Fisher Information (QFI) and entanglement entropy while significantly reducing gate count and depth. We use Double Deep Q-Network (DDQN) enhanced with entanglement-aware attention and a custom reward function that balances metrological performance and circuit efficiency. Quantum circuit optimization is fundamentally a sequential decision-making problem with a combinatorially large action space and a non-differentiable reward landscape. Classical optimization methods, such as gradient descent or rule-based heuristics, either fail to scale or require access to gradients which are unreliable or unavailable in noisy quantum environments.

In our context, RL enables a policy to be learned over gate-level transformations—such as inserting, removing, or reordering gates—based on reward signals tied to entanglement entropy, circuit depth, and Quantum Fisher Information (QFI). Rather than optimizing fixed parameters, the RL agent learns how to restructure circuits themselves, making it far more flexible than static optimization algorithms. The process begins with an initial quantum circuit, which is then iteratively modified by the DDQN agent. The agent learns from a deep Convolution network where it chooses between several circuit transformations or to generate another logically equivalent circuit by logically transforming quantum circuit to Tensor nodes to efficiently simulate the larger quantum state. The process is repeated to achieve the best Reward.

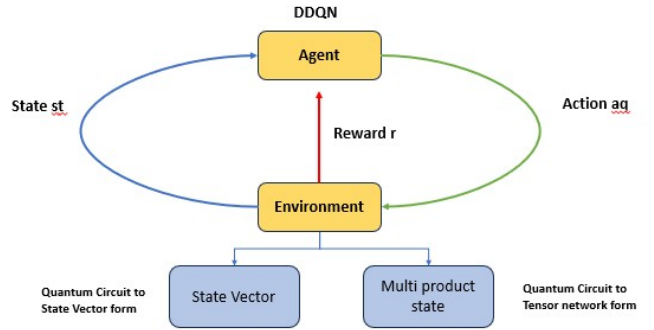

Fig. 2. Reinforcement learning framework with traditional statevector method and Tensor network method.

Standard Deep Q-Networks (DQN) are prone to overestimation bias, which becomes a problem in environments like ours where rewards are sparse and noisy (especially from stochastic QFI evaluations). DDQN mitigates this by decoupling action selection and evaluation using separate networks where Online network chooses the action and Target network evaluates its value.

The environment is modeled as a Markov Decision Process (MDP) where the state is represented by the current quantum circuit configuration. The DDQN algorithm is used to learn the optimal policy for circuit modification. The Q-value update follows the equation:

$$Q(s,a) \leftarrow Q(s,a) + \alpha[r + \gamma\, Q'(s', \text{argmax}_a\, Q(s',a)) - Q(s,a)] \quad (1)$$

where Q and Q' are the main and target networks respectively, α is the learning rate, γ is the discount factor, r is the reward, and s and s' are the current and next states. In addition to these layers, we also added experience reply to break correlations in the observation sequence and smooth over changes in the data distribution, experiences (s, a, r, s') are stored in a replay buffer and sampled randomly for training. A separate convolution layer to focus on entanglement and critic network layer [10], [11] to evaluate the value of state-action pair to help the agent to differentiate the action of benefits during the learning process. The environmental actions correspond to various circuit modifications such as adding gates, removing gates, entanglement injections, entanglement boosting and swapping gate positions. The RL agent learns to navigate this action space with a set of gates like Hadamard to create superposition that internally allows the qubit to exist in multiple states simultaneously to improvise the sensitivity of quantum sensor circuits. Entanglement gates such as CNOT, CZ, SWAP gates were used to create the entangled state along with few rotational gates like RX and RZ with parameter angle theta to rotate around x-axis and z-axis of the blotch sphere respectively, to maximize the precision of the entangled quantum sensor circuit.

Traditional quantum simulation techniques rely on full statevector or density matrix representations. However, these approaches scale poorly with the number of qubits requiring $O(2^n)$ and $O(4^n)$ memory and time complexity respectively making them impractical for simulating circuits beyond 20 qubits. To overcome this, we adopt the MPS formalism, where a quantum state $|\psi\rangle|$ is decomposed into a chain of tensors, with each tensor representing local entanglement between adjacent qubits. Formally, the quantum state $|\psi\rangle \in H_1 \otimes H_2 \cdots \otimes H_n$ is expressed as

$$|\psi\rangle = \sum_{i_1,\ldots,i_n} A_{i_1}^{[1]} A_{i_2}^{[2]} \cdots A_{i_n}^{[n]} |i_1 i_2 \cdots i_n\rangle \qquad (2)$$

where each $A^{[k]}$ is a three-dimensional tensor whose internal bond dimension $\chi$ determines how much entanglement can be captured between sites. This reduces simulation cost to $O(n\chi^2)$, which remains tractable when entanglement is moderate or locally structured.

A reward function in reinforcement learning is a crucial component that guides an agent's behavior by assigning a numerical value to its actions within a given state. This value, representing a reward or penalty, influences the agent's decision-making process as it strives to maximize its cumulative reward over time. As in our scenario, single reward function is not sufficient to meet the realistic challenges to achieve optimization with high precision. To address these challenges, we have incorporated multi reward function with different weights to balance the agent goal while optimizing the entanglement and circuit complexity reduction. The reward function is a weighted sum of improvements in QFI, depth reduction, entanglement enhancement, and gate count reduction:

$$R = w_1 * \Delta QFI + w_2 * \Delta Depth + w_3 * \Delta Entropy + w_4 * \Delta Gates \qquad (3)$$

where $\Delta$ represents the change in each metric, and w1, w2, w3, w4 are weight parameters.

To quantify and optimize entanglement, we compute the von Neumann entropy across every MPS bond using the Schmidt decomposition. The entanglement entropy across a bipartition is defined as

$$S = -\sum_i \lambda_i^2 \log_2 \lambda_i^2 \qquad (4)$$

, where $\lambda_i$ are the Schmidt coefficients obtained from the MPS. Entropy values are normalized to the range [0, 1] and continuously tracked by the agent. During training, entropy is used both as a reward signal and as a trigger for injecting entanglement if it drops below a threshold, initially set to 0.7 and dynamically adjusted per episode.

In parallel, the framework evaluates the Quantum Fisher Information (QFI) of the circuit, which serves as a key metric for the sensitivity of quantum sensors. For parametrized circuits, QFI is approximated using a parameter-shift technique combined with measurement statistics. Each parameter $\theta$ is shifted by $\pm\Delta$ where $\Delta=\pi/2$ and the corresponding probabilities $P^i\pm$ of measurement outcomes i are collected over 5000 shots. The QFI for a parameter is computed as

$$QFI(\theta) = \sum_i 4(P_+^i - P_-^i)^2 / (P_+^i + P_-^i) \qquad (5)$$

and averaged across all parameters.

Depth ratio is used as one of the rewards to measure the Depth Change and focuses on minimization of Circuit Depth. This reward helps in reducing the complexity of overall circuits by keeping the state of entanglement unchanged.

$$Depth\ Ratio = D_{in} - D_{out} / D_{in} \qquad (6)$$

where $D_{in}$ is the depth of input and $D_{out}$ is the depth of output circuit.

Gate ratio is used as one of the rewards to measure the gate counts change and focuses on minimization of gate counts. This reward helps to reduce the noise of overall circuits by cancelling noisy and unwanted gates.

$$Gate\ Ratio = G_{in} - G_{out} / G_{in} \qquad (7)$$

where $G_{in}$ is the gates of input and $G_{out}$ is the gates of output circuit.

During training, the agent selects actions using an epsilon-greedy policy, with the epsilon value decaying from 1.0 to 0.01 over episodes. The agent's experience is stored in a memory buffer, and entangling actions are prioritized during replay to encourage the formation of high-quality entanglement structures.

This methodology not only enables the scalable optimization of large quantum sensor circuits but also provides a robust learning-based alternative to heuristic or gradient-based techniques. The RL agent adapts dynamically to circuit structure and noise, generalizing across architectures and entanglement patterns while maintaining a strong balance between quantum utility and practical implement ability.

## V. EXPERIMENTS AND RESULTS

The experiment was designed to evaluate the effectiveness of a Reinforcement Learning framework, enhanced with tensor network simulation, in optimizing entanglement within large-scale quantum circuits. A custom Quantum Circuit Environment was created using OpenAI Gym, simulating a quantum circuit with a specified number of qubits from 5 to 60 and a maximum number of gates from 15 to 120 respectively. The agent was trained for 5-50 episodes, with each episode starting from one of the loaded initial circuits. To provide more flexible action space we have added many actions like the agent could perform various actions including adding different types of quantum sensing gates (H, CNOT, RX, RZ, CZ, SWAP), removing gates, swapping gate positions, gate cancellation, injection, boosting and replacing gates. The combination of gates for creation of superposition, entanglement and phase shift with amplitude amplification clearly differentiate the quantum sensor circuits from regular quantum circuits that are specifically designed for quantum sensing networks. Also, by keeping DDQN as the base network layer we have experimented with two additional network layers, one is the Critic network layer to evaluate the agent actions and entanglement focused network layer to understand the behavior of entangled gates by taking previous entanglement features as an input. The setup of the environment and hyperparameters used during the training are given in detail below.

| Elements | Value | Description |
|---|---|---|
| Packages used | Qiskit, TensorFlow, keras, NumPy, gym, matplotlib, sklearn, pandas | All the classical and quantum packages that are used to process the proposed DDQN approach |
| Training episodes | 5-50 | The size of the training episodes to train the agent |
| Qubits | 5-60 | The size of the Qubits considered for the optimization |
| Quantum simulator | AER, state vector, MPS | Simulators used for simulation |
| Memory size Batch size | 2000 128 | Maximum number of experiences sampled from the reply buffer |
| Discount factor(gamma) | 0.95 | Used in Q-value update |
| Epsilon Decay | 0.999 | The rate at which the exploration rate (ε) is decreased |
| Entanglement threshold | 0.7 | The minimum entanglement level required for the circuit. |

TABLE 1. The setup and hyperparameters of the environment

Initially we started the experiment by training the agent with 5 qubit circuit to analyze the state of DDQN architecture. As the number of gates and qubits are minimal for computation the results are bit less due to minimal opportunities for exploration.Moving forward we trained the agent for 8,15,20,25,35,50 and 60 qubits with higher number of gates upto 120 to provide more action space for exploration and exploitation.The results were quite satisfactory and we are able to achieve the goal of balancing the multi reward functions by maximizing the QFI and entanglement of the circuit by minimizing the complexity of circuit with depth and gate counts reduction.Before applying different optimization techniques we were able to optimize the entanglement layout with maximum QFI and entropy of average 0.80 to 0.90 with average depth reduction of 3 to 16 % and max of 86% and gate counts reduction of 6 to 29% on an average and 94% of maximum .

But as our main goal is to achieve the high entangled circuit we have adopted few optimization techniques to focus on the area of entanglement within the quantum sensor circuit. To stabilize and enhance the learning process, an adaptive learning rate mechanism is applied. The learning rate is adjusted in real-time based on training progress using an exponentially decaying schedule combined with performance-based decay triggers.

This is optimized using the Adam optimizer, which dynamically tunes parameter updates by tracking first and second moments of the gradients, offering fast convergence and using qiskit's noise model with depolarizing and thermal relaxation errors, our approach maintained up to 27% average gate reduction and QFI values >0.8 across 50 test circuits.. To overcome the scalability limitations inherent in traditional quantum simulators, the experiment utilizes a Matrix Product State (MPS) based tensor network backend. This allows efficient simulation of circuits with qubit counts extending beyond 20, a regime where full statevector or density matrix methods become computationally infeasible due to exponential growth in memory and time complexity. To prove the ability and strength of tensor network we have compared the results of state of art [7] and statevector with MPS [15], [16] based results up to 20 qubits. As per the expectations the results of Tensor approach is like the results of statevector along with reduction in time and memory consumption by 2 times.

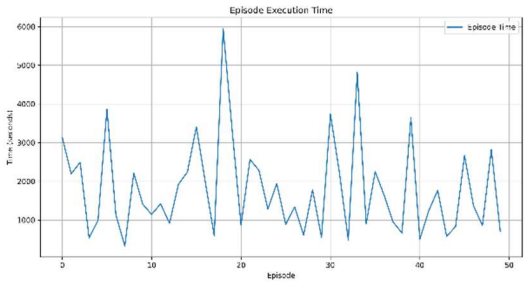

Fig. 3. The execution time for simulating a 20-qubit quantum circuit using the statevector method was recorded at approximately **90122 seconds**, highlighting the computational cost associated with large-scale entanglement-aware simulation.

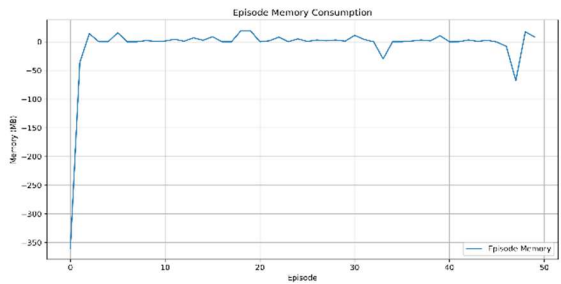

Fig. 4. Total memory consumption for simulating a 20-qubit quantum circuit using the state vector method was recorded at 346MB.

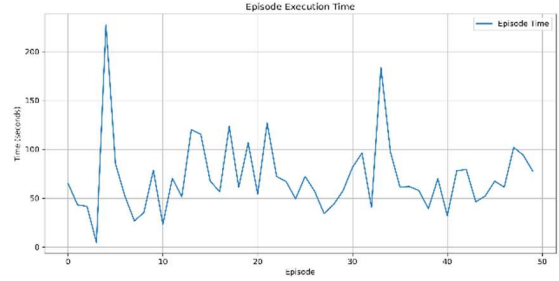

Fig. 5. The execution time for simulating a 20-qubit quantum circuit using the tensor-based MPS method was recorded at approximately **3,632 seconds**.

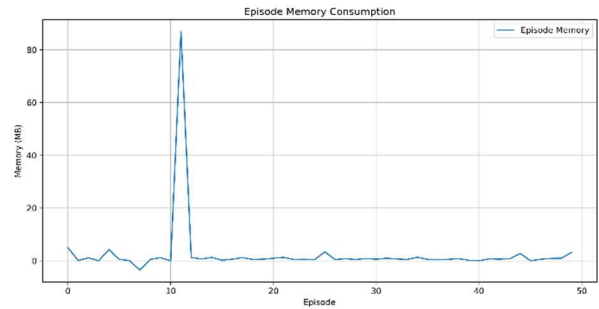

Fig. 6. Total memory consumption for simulating a 20-qubit quantum circuit using the tensor-based MPS method was recorded at 129MB.

A dynamic entanglement threshold is maintained to ensure the optimization process does not reduce quantum resources below a functional level. This threshold is not static; it adapts based on observed circuit behavior, allowing for flexibility during simplification or pruning.

To ensure that the optimization process reflects realistic quantum hardware conditions, we integrated a comprehensive noise model into the simulation environment using Qiskit's NoiseModel class. The simulation is executed on the AerSimulator backend, specifically utilizing the matrix_product_state (MPS) method to support efficient and scalable simulation of large quantum circuits. The noise model incorporates depolarizing errors for both measurement operations and quantum gates, as well as thermal relaxation noise representing T1 and T2 decoherence processes. Specifically, the experimental settings uses Measurement error rate: pmeas=0.02,Single-qubit gate depolarizing error: p1q=0.01,Two-qubit gate depolarizing error: p2q=0.03,T1 time constant: 50 µs,T2 time constant: 70 µs

These realistic noise parameters emulate the behavior of near-term quantum devices and allow the reinforcement learning (RL) agent to generate robust circuit architectures that can withstand practical error conditions. The MPS method enables simulations to scale polynomially under limited entanglement conditions. This allowed us to simulate and optimize quantum circuits with up to 35 qubits and 120 gates on a local machine with 8 GB RAM and 1 TB storage, without requiring quantum hardware or cloud-based resources. To ensure smoother execution and faster training cycles, the optimization and simulation processes were run on Google Colab equipped with an A100 GPU, significantly accelerating both circuit evaluations and model training compared to standard CPU-based environments.

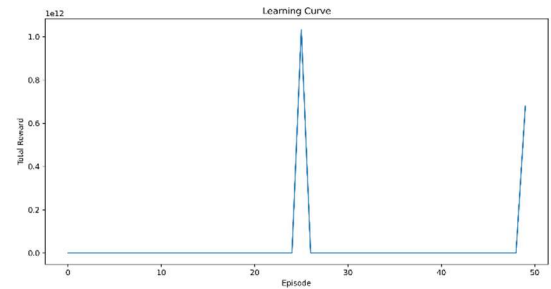

Fig. 7. Total reward curve of 20 qubit circuits with statevector method is compared with the reward curve of Tensor method below.

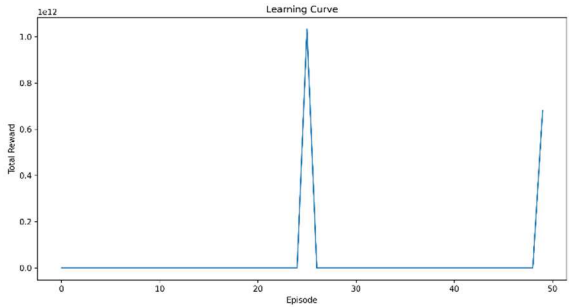

Fig. 8. Total reward curve of 20 qubit circuits with tensor method, proves the stability of model while running irrespective of the simulation method.

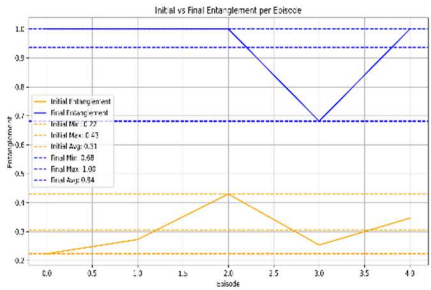 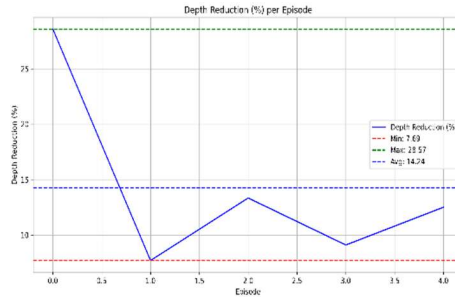 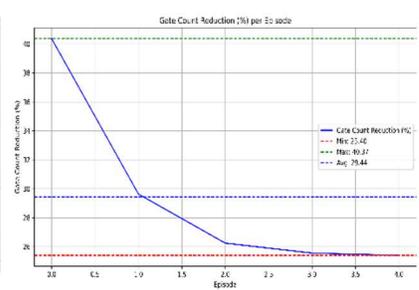

9a  9b  9c

Figure 9a presents a comparison between the initial and final entanglement entropy values for a 25-qubit quantum circuit optimized using the MPS-based simulation approach. Figure 9b shows that the RL optimizer reduced depth by up to 16% for 25-qubit circuits, demonstrating its ability to minimize complexity without degrading entanglement. Figure 9c shows the gate count reduction curve, demonstrating the agent's ability to minimize redundant or unnecessary operations while preserving entanglement and functionality—all for a 25-qubit circuit simulated via the tensor network method.

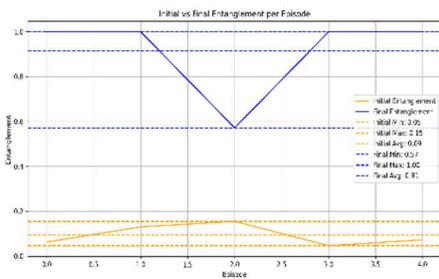 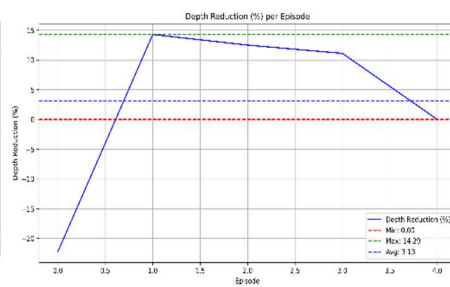 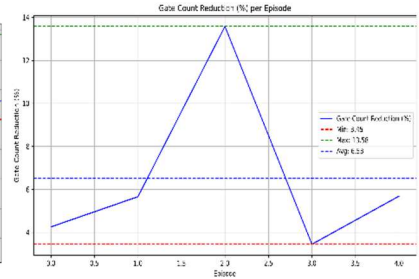

10a  10b  10c

Figure 10a presents a comparison between the initial and final entanglement entropy values for a 35-qubit quantum circuit optimized using the MPS-based simulation approach. Figure 10b illustrates the percentage reduction in circuit depth achieved through the reinforcement learning optimization process. Figure 10c shows the gate count reduction curve, demonstrating the agent's ability to minimize redundant or unnecessary operations while preserving entanglement and functionality—all for a 25-qubit circuit simulated via the tensor network method.

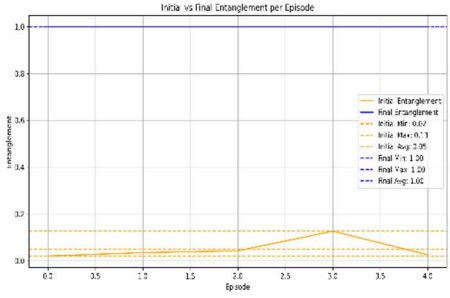 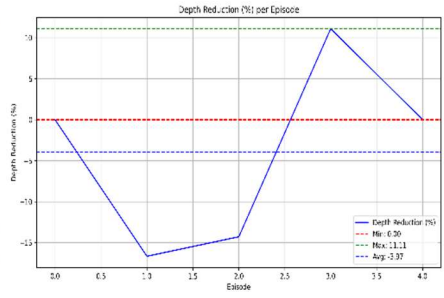 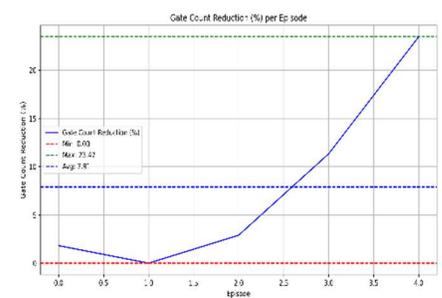

11a  11b  11c

Figure 11a presents a comparison between the initial and final entanglement entropy values for a 50-qubit quantum circuit optimized using the MPS-based simulation approach. Figure 11b illustrates the percentage reduction in circuit depth achieved through the reinforcement learning optimization process. Figure 11c shows the gate count reduction curve, demonstrating the agent's ability to minimize redundant or unnecessary operations while preserving entanglement and functionality—all for a 25-qubit circuit simulated via the tensor network method.

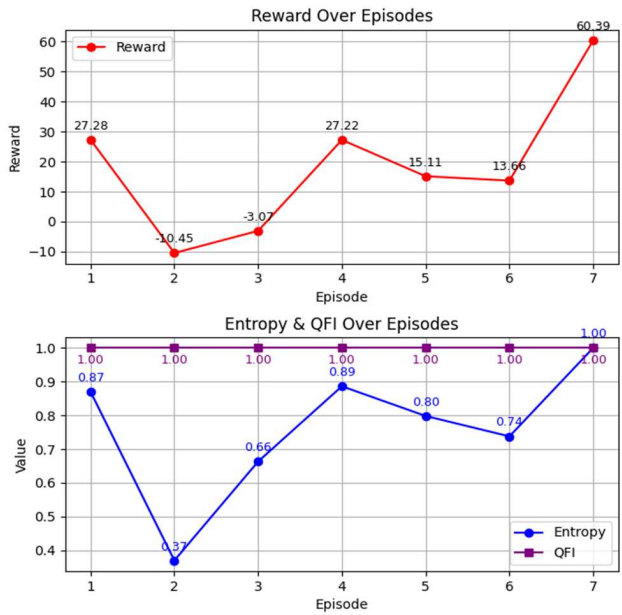
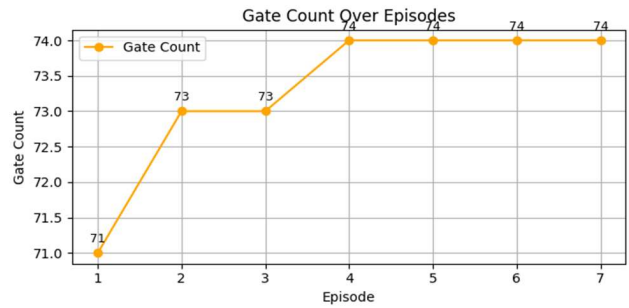
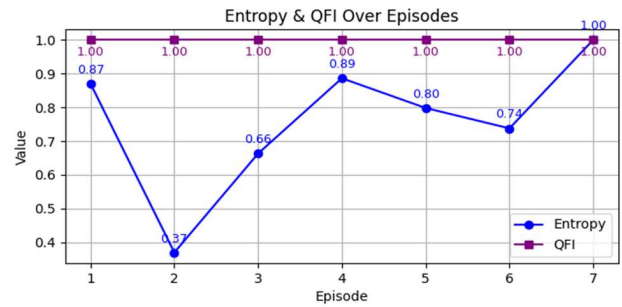

Figure 12. Visualization of 60 qubit circuits optimized over simulation of 7 episodes with achievement of maximum QFI and Entanglement from 0.59 to 1.0 depth reduction from 7 to 5 , gate counts reduction from 74 to 68 which proves the ability of model to maximize the precision of quantum sensor circuits by maximizing QFI and entanglement along with the minimization of circuit complexity by depth and gate counts reduction.

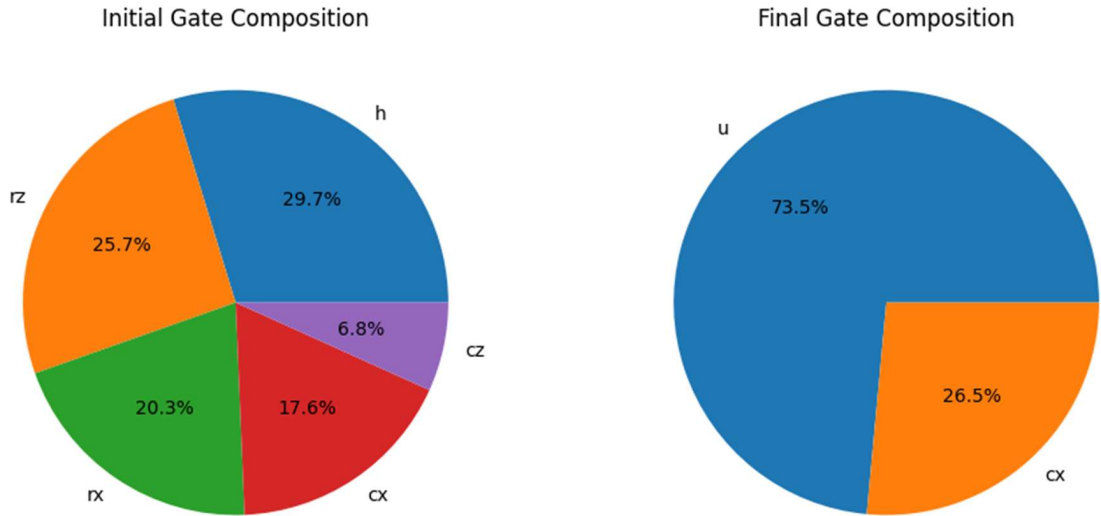

Figure 13. The pie charts illustrate the relative proportions of different quantum gates used in the circuit before and after optimization. Initially, the circuit comprised a mix of H, RZ, RX, CX, and CZ gates, with H gates being the most frequent (29.7%). After optimization, the circuit is reduced to primarily universal single-qubit gates (U, 73.5%) and fewer two-qubit CX gates (26.5%), indicating gate simplification and consolidation for improved circuit efficiency.

| Q(n) | Max gates | QFI | | | Entropy | | | Depth Reduction % | | | Gates Reduction % | | |
|---|---|---|---|---|---|---|---|---|---|---|---|---|---|
| | | i | ii | iii | i | ii | iii | i | ii | iii | i | ii | iii |
| 5 | 30 | **1.0** | **1.0** | **1.0** | **0.86** | **0.94** | **0.92** | **16.48** | **5.08** | **4.97** | **13.94** | **11.47** | **11.55** |
| 8 | 60 | **1.0** | **1.0** | **1.0** | **0.92** | **0.96** | **0.93** | **8.29** | **10.04** | **9.7** | **4.68** | **17** | **16.84** |
| 15 | 90 | **1.0** | **1.0** | **1.0** | **0.56** | **0.59** | **0.57** | **-2.33** | **10.14** | **16.27** | **2.11** | **16.12** | **16.80** |
| 20 | 100 | **1.0** | **1.0** | **1.0** | **0.85** | **0.82** | **0.88** | **-3.89** | **8.34** | **6.19** | **1.72** | **9.91** | **11.3** |
| 25 | 100 | **NA** | **NA** | **1.0** | **NA** | **NA** | **0.94** | **NA** | **NA** | **14.24** | **NA** | **NA** | **29.44** |
| 35 | 120 | **NA** | **NA** | **1.0** | **NA** | **NA** | **0.91** | **NA** | **NA** | **3.13** | **NA** | **NA** | **6.53** |
| 50 | 120 | **NA** | **NA** | **1.0** | **NA** | **NA** | **1.0** | **NA** | **NA** | **-3.97** | **NA** | **NA** | **7.91** |
| 60 | 120 | **NA** | **NA** | **1.0** | **NA** | **NA** | **1.0** | **NA** | **NA** | **28.57** | **NA** | **NA** | **8.11** |

TABLE 2. The Results table of 5 to 60 Qubits Quantum circuits simulation using Tensor approach and Qiskit compiler with multi metrics QFI, Entropy, Depth Reduction and Gate counts Reduction compared to the baseline work [7] i) Result values estimated with state-of-the-art[7]. ii) Result Values estimated with traditional statevector method. iii) Result values estimated with optimization techniques and tensor method to achieve the scalability beyond 20 qubits. The Results clearly state that the model can achieve highest QFI and Entanglement Entropy of average 0.80 to 0.92 and max of 1.0 along with the average depth and gate counts reduction by 25% and maximum by 90%. The NA values in the table indicates that the simulation or estimation is impossible for qubits beyond 20 with traditional statevector method.

## VI. Conclusion and future work

This research successfully demonstrates the effectiveness of a reinforcement learning-based framework for optimizing quantum sensor circuits, particularly with a large number of qubits. By combining Double Deep Q-Network (DDQN) reinforcement learning with a tensor network simulation backend specifically the Matrix Product State (MPS) representation we address both the optimization complexity and scalability limitations inherent in quantum circuit design. The experiments effectively enhance the sensitivity of quantum sensor circuits by optimizing the distribution and quality of entanglement—a core resource in quantum metrology.

The primary contribution of this work lies in its ability to maximize Quantum Fisher Information (QFI) and entanglement entropy, both of which are direct indicators of a quantum circuit's potential for high-precision measurement. Across various 5-60 qubit configurations, the framework consistently achieved QFI and entropy values in the range of 0.8 to 1.0, reflecting near-optimal sensitivity. This is accomplished without sacrificing efficiency of circuit depth and gate counts were reduced by up to 90%, demonstrating the model's ability to maintain or even improve performance while simplifying the circuit structure.

The hybrid architecture integrating reinforcement learning with scalable tensor network simulation proves to be highly effective in navigating the complex multi-objective landscape of quantum circuit optimization. It enables the model to make intelligent decisions about circuit restructuring while remaining resilient to noise and capable of generalizing to larger and more complex circuits.

Ultimately, this framework not only advances the current state of quantum circuit optimization but also lays the groundwork for practical deployment of optimized quantum sensor networks.

While the current implementation successfully optimizes circuits up to 60 qubits using the Matrix Product State (MPS) method, future work will focus on extending this framework to simulate and optimize circuits with 100 or more qubits. This will involve enhancing the tensor network backend, potentially integrating more advanced tensor formats such as Tree Tensor Networks (TTN) or Projected Entangled Pair States (PEPS), which are better suited for capturing entanglement in more complex, higher-dimensional systems. Further automation of gate sequence restructuring will be a critical step. This includes integrating compilation-aware optimization techniques that not only reduce gate count and circuit depth but also align with hardware-specific constraints. To address the impact of noise and imperfections in realistic quantum systems, future work will incorporate error mitigation strategies directly into the reinforcement learning framework. Techniques such as zero-noise extrapolation, probabilistic error cancellation, and readout error mitigation will be considered. Given the computational demand of simulating and training large quantum circuits, future iterations will parallelize key computations using multi-threaded execution and GPU acceleration.

Finally, to validate the practicality of the proposed methods, the framework will be extended to execute on real quantum hardware platforms such as IBM Quantum. This transition from simulation to hardware will highlight performance gaps, calibration challenges, and device-specific constraints, and offer insights into how reinforcement learning and tensor methods can adapt to time and space limitations inherent to physical quantum systems.